\documentclass{article}
\usepackage{spconf,amsmath,graphicx}
\usepackage{booktabs}
\usepackage{xcolor}
\usepackage{hyperref}
\hypersetup{hidelinks}
\title{Frozen Site-Probe Accuracy Is Not a Standalone Harmonization Score:\\
An Audit of Evaluation Practice for Multi-Site MRI}

\name{
\parbox{\linewidth}{\centering
Muhammad Muqsit Islam$^{1,2,\star,\dagger}$\thanks{$^\star$Corresponding author}\thanks{$^\dagger$ Equal contribution}
\qquad
Gloria García Cuenco$^{1,2,\dagger}$ \\
\qquad José M. Martínez$^{3}$ \qquad
Pierrick Coupé$^{4}$ \qquad \\
}}

\address{$^{1}$ Universidad Autónoma de Madrid, Spain \qquad $^{2}$ Université de Bordeaux, France \\
$^{3}$ Video Processing and Understanding Lab, Universidad Autónoma de Madrid, Spain \\
$^{4}$ Université de Bordeaux, CNRS, Bordeaux INP, LaBRI, UMR 5800, F-33400 Talence, France}

\begin{document}
\raggedbottom
\maketitle

\begin{abstract}
Image-level MRI harmonization is often evaluated with a frozen site probe, a site classifier trained on raw scans and applied unchanged to harmonized ones: a drop in its accuracy is read as removal of site effects. We audit this practice on ABIDE~I (1,112 T1-weighted scans, 17 sites) with ten output sets from seven harmonization methods and controls for chance, head shape, intensity, preprocessing and target alignment. We find that this verdict depends on how the probe is trained, even among probes that classify raw scans well: across random seeds, its balanced accuracy on the same CycleGAN outputs ranges from 0.13 to 0.59 (chance: 0.06). Low frozen-probe accuracy can also arise without harmonization: a Gaussian blur, which increases the intensity distance to the target site, lowers this accuracy to chance. Moreover, where the frozen probe fails, probes trained on the harmonized scans still recover the site, even from intensity histograms alone: on HACA3 outputs, the frozen probe scores 0.25, a probe trained on those outputs 0.75, and one using only their intensity histograms 0.71. Low frozen-probe accuracy therefore does not show that site information has been removed. We recommend reporting several probes trained to convergence, probes trained on harmonized scans, and target alignment.
\end{abstract}

\begin{keywords}
MRI harmonization, evaluation, site classification, domain shift, reproducibility
\end{keywords}

\section{Introduction}
\label{sec:intro}
Differences between scanners and protocols confound pooled analyses of multi-site MRI and degrade models on unseen sites~\cite{fortin2018harmonization,glocker2019multisite}. Image-level harmonization addresses this by translating scans toward a common appearance. Whether it succeeds is hard to verify: direct checks need travelling subjects~\cite{tanaka2021srpbs} or paired acquisitions~\cite{dewey2019deepharmony}, which exist for only a few scanners, so large retrospective cohorts are evaluated indirectly~\cite{hu2023harmonization}.

A common indirect check is a \emph{site probe}, a classifier trained to predict the acquisition site, a task that is easy on raw MRI~\cite{glocker2019multisite}. If its accuracy drops on harmonized images, site effects are considered removed~\cite{wu2025dlest,wu2026hcld}. Some studies retrain the probe on harmonized images~\cite{wu2025dlest} or image features~\cite{wu2026hcld,dinsdale2021unlearning}; others train it once on original images and apply it unchanged, which we call a \emph{frozen} probe. PRISM~\cite{galada2025prism}, for example, reports such a drop, with target-site predictions and segmentation checks, as evidence of harmonization. Probe accuracy is usually reported alongside input similarity measures, such as PSNR.

These checks can mislead. A frozen probe may fail simply because harmonized images look unfamiliar, not because site information is gone. Removing an attribute until a fixed classifier fails does not prevent a new classifier from recovering it~\cite{elazar2018adversarial}, and site can be decoded even from the features of randomly initialized networks~\cite{rahbar2026fingerprints}. Input similarity, in turn, can favor little or no change.

Classifiers retrained after harmonization still predict site for ComBat~\cite{marzi2024efficacy}, histogram matching~\cite{souza2023shortcuts} and HACA3~\cite{zurita2026psyshared}. Less clear is how much different probe protocols disagree on the same harmonized images. We audit this on ten output sets from seven methods, starting from a frozen probe: how much its verdict depends on training, whether low accuracy persists with output-trained probes, and what controls for head shape, intensity and target alignment add. We do not rank harmonizers; we derive a reporting checklist (Sec.~\ref{sec:discussion}).\footnote{Supplementary material (code, training settings, protocol, additional statistics and per-subject results): \url{https://github.com/muqsitamir/HarmonIt}.}

\section{Data and Methods}
\label{sec:data}
\textbf{Dataset.} We use raw T1-weighted scans of 1,112 ABIDE~I subjects from 17 sites~\cite{dimartino2014abide}, split by subject and stratified by site into 890 training, 113 validation and 109 test subjects. Each volume is clipped to its 1st--99th intensity percentiles and scaled to $[0,1]$. We keep one axial slice per subject, the one with the largest head area, cropped to the head and resized to $256\times256$; every method is evaluated on the same slices. NYU, the largest site, is the harmonization target, so we evaluate the 90 test subjects from the 16 other sites. Site labels are coarser than acquisitions: UM, UCLA and Leuven each merge two released sub-samples, and 11 of the 17 sites mix voxel sizes or matrices.

\textbf{Harmonization methods.} We evaluate outputs of six methods implemented by us: NeuroCombat~\cite{fortin2018harmonization,johnson2007combat}, applied per pixel and fit on the test subjects themselves; histogram matching~\cite{nyul2000standardization} to a reference pooled over all training sites; CycleGAN~\cite{zhu2017cyclegan}; StarGAN~\cite{choi2018stargan} in an aggressive and a conservative setting; a DLEST-style model~\cite{wu2025dlest} without its energy-based sampler, after 1,000 and 1,500 training steps; and conditional diffusion translation~\cite{ho2020ddpm,meng2022sdedit}, sampled twice. HACA3~\cite{zuo2023haca3} is used with its authors' pretrained weights and an NYU target image; its required bias correction and registration to MNI space are undone afterwards, so its outputs align with the original slices. As a control, we also blur the raw test slices with a Gaussian filter, which changes the images without moving them toward NYU.

\begin{table*}[t]
\centering
\caption{Results on the 90 source test subjects (means). Arrows give the direction usually read as better. Change: similarity to the input (PSNR, XCorr) and distance between input and output intensities ($W$). Target alignment: change in Wasserstein distance ($\Delta W_{\mathrm{NYU}}$) and KL divergence ($\Delta$KL$_{\mathrm{NYU}}$) to NYU intensities. Site BA: balanced accuracy of probes trained on unharmonized images and tested on each row; frozen probe with 95\% confidence interval and share of outputs it labels NYU; reseeded and converged: mean (min--max) across five seeds; blur: frozen probe only. Probes trained on the outputs: Fig.~\ref{fig:verdicts}. $^\dagger$Fit on the test subjects. $^\ddagger$Reference pooled over all sites, not just NYU. $^\S$Renders CSF near zero, which inflates $\Delta$KL$_{\mathrm{NYU}}$.}
\label{tab:main}
\vspace{2pt}
\scriptsize
\setlength{\tabcolsep}{3.5pt}
\begin{tabular}{@{}lccccccccc@{}}
\toprule
 & \multicolumn{3}{c}{Change} & \multicolumn{2}{c}{Target alignment} & \multicolumn{4}{c}{Site BA: probes trained on unharmonized images, tested on each row} \\
\cmidrule(lr){2-4}\cmidrule(lr){5-6}\cmidrule(l){7-10}
Method & PSNR (dB)$\uparrow$ & XCorr$\uparrow$ & $W$ & $\Delta W_{\mathrm{NYU}}\!\downarrow$ & $\Delta$KL$_{\mathrm{NYU}}\!\downarrow$ & Frozen probe & Pred. NYU (\%)$\uparrow$ & Reseeded & Converged \\
\midrule
Raw input & -- & -- & 0 & 0 & 0 & 0.95 & 0 & 0.90 (0.81--0.98) & 1.00 (1.00--1.00) \\
NeuroCombat$^\dagger$ & 20.0 & 0.940 & 0.055 & +0.003 & +0.57 & 0.27 [0.21, 0.32] & 2 & 0.33 (0.26--0.37) & 0.34 (0.28--0.42) \\
Histogram matching$^\ddagger$ & 23.0 & 0.982 & 0.054 & -0.053 & -0.22 & 0.44 [0.37, 0.52] & 10 & 0.44 (0.27--0.59) & 0.61 (0.55--0.68) \\
CycleGAN & 21.3 & 0.977 & 0.048 & -0.036 & -0.06 & 0.31 [0.25, 0.37] & 52 & 0.28 (0.13--0.59) & 0.49 (0.44--0.54) \\
StarGAN (aggr.) & 16.0 & 0.896 & 0.071 & -0.019 & +0.02 & 0.21 [0.15, 0.27] & 56 & 0.35 (0.18--0.54) & 0.70 (0.65--0.81) \\
StarGAN (cons.) & 26.9 & 0.991 & 0.021 & -0.021 & -0.03 & 0.83 [0.79, 0.86] & 1 & 0.78 (0.66--0.89) & 0.98 (0.94--0.99) \\
DLEST-style 1000 & 28.4 & 0.993 & 0.017 & -0.009 & +0.00 & 0.71 [0.65, 0.77] & 6 & 0.84 (0.76--0.89) & 0.99 (0.96--1.00) \\
DLEST-style 1500 & 26.6 & 0.993 & 0.024 & -0.020 & -0.06 & 0.67 [0.61, 0.72] & 10 & 0.73 (0.65--0.82) & 0.96 (0.95--0.97) \\
Diffusion, draw 1 & 21.7 & 0.948 & 0.008 & -0.001 & +0.53 & 0.37 [0.32, 0.42] & 36 & 0.42 (0.24--0.63) & 0.57 (0.52--0.59) \\
Diffusion, draw 2 & 21.7 & 0.948 & 0.008 & -0.002 & +0.51 & 0.35 [0.29, 0.40] & 38 & 0.42 (0.30--0.62) & 0.56 (0.50--0.59) \\
HACA3 (pretrained)$^\S$ & 18.9 & 0.954 & 0.062 & -0.016 & +1.98 & 0.25 [0.18, 0.32] & 40 & 0.29 (0.18--0.37) & 0.28 (0.23--0.33) \\
\midrule
Blur control, $\sigma{=}2$ & 24.7 & 0.974 & 0.018 & +0.016 & +0.27 & 0.07 [0.06, 0.10] & 0 & -- & -- \\
\bottomrule
\end{tabular}

\end{table*}

\begin{figure}[t]
\centering
\includegraphics[width=\linewidth]{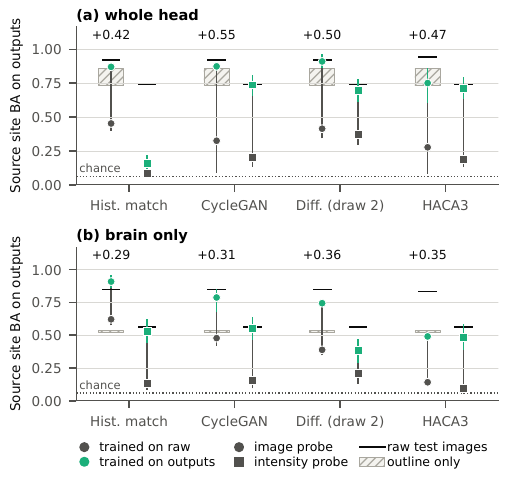}
\caption{Is site still decodable? Raw-trained (grey) and output-trained (aqua) probes tested on harmonized (a) whole-head and (b) brain-only slices; diffusion uses draw 2. Circles: image probes (mean and range across three seeds); squares: intensity-histogram probes (95\% confidence intervals). Numbers: absolute BA gain from output training. Black bars: raw-trained probes tested on raw slices; hatched boxes: outline-only probes. Complements Table~\ref{tab:main}, which reports only probes trained on unharmonized images.}
\label{fig:verdicts}
\end{figure}

\begin{figure*}[t]
\centering
\includegraphics[width=\textwidth]{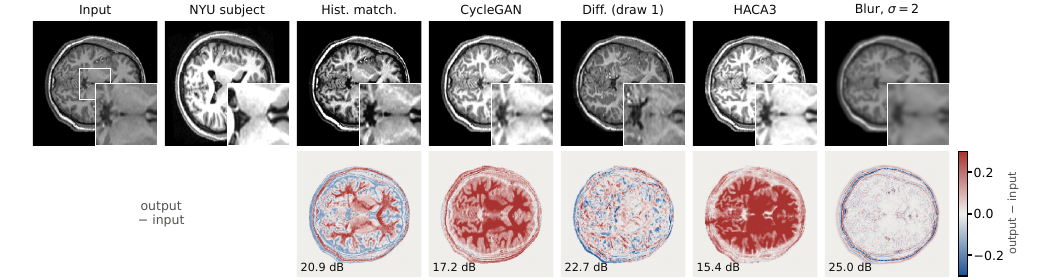}
\caption{An illustrative source subject and an NYU test reference selected by whole-head Dice $\times$ boxed-region correlation. Images share the input's display window; insets magnify the box $2\times$. Bottom: output $-$ input, with PSNR.}
\label{fig:qualitative}
\end{figure*}

\textbf{Image change and target alignment.} We compare each output with its input using PSNR and normalized cross-correlation (XCorr), and measure how much its intensity distribution changed with the Wasserstein-1 distance $W$. To check whether outputs moved toward the target, we compare their intensity distribution with that of the 147 NYU training subjects (Wasserstein distance and KL divergence); $\Delta W_{\mathrm{NYU}}$ and $\Delta$KL$_{\mathrm{NYU}}$ are the output's distances minus the input's, so negative values mean closer to NYU.

\textbf{Site probes.} Image probes are ResNet-18 classifiers~\cite{he2016resnet} that predict the site of a slice; we report balanced accuracy (BA) over the 16 source sites. We compare five families of probes. The first four draw one random axial slice per raw training volume at each step (Table~\ref{tab:main}); the fifth uses fixed raw or harmonized training slices (Fig.~\ref{fig:verdicts}). All are tested on fixed slices.
\begin{list}{}{\setlength{\leftmargin}{1em}\setlength{\itemindent}{0pt}\setlength{\topsep}{2pt}\setlength{\itemsep}{1pt}\setlength{\parsep}{0pt}}
\item \emph{Frozen}: one probe trained once on raw images with a short schedule (500 steps, best validation epoch; raw BA 0.95) and applied unchanged to every output.
\item \emph{Reseeded}: the frozen probe's training recipe rerun with five random seeds, one of them its own.
\item \emph{Converged}: five probes trained four times longer, until validation accuracy is stable.
\item \emph{Shuffled labels}: probes trained on permuted labels, giving the chance level actually reached (0.03--0.08; uniform guessing gives 0.06).
\item \emph{Slice probes}: trained on one fixed slice per training subject, using either the raw slices or a method's harmonized slices (three seeds each), and both tested on that method's harmonized test slices. If the probe trained on harmonized slices scores much higher, site information is still in the images. Control versions see only the head or brain outline, only intensity histograms, or only the brain (HD-BET masks~\cite{isensee2019hdbet}).
\end{list}
Confidence intervals are 95\% percentile bootstrap intervals from 2,000 resamples of the test subjects, stratified by site and shared across probes; they reflect these subjects and trained probes, not uncertainty across retraining or new cohorts.

\section{Results}
\label{sec:results}
\textbf{Probes see more than scanner appearance.} The probes in Table~\ref{tab:main} achieved BA 0.81--1.00 on raw test images. Head outlines alone already gave 0.73--0.86, whereas age, sex and diagnosis gave only 0.22, so site is visible in head geometry, not only in image contrast.

\textbf{The verdict depends on how the probe is trained.} Although the frozen probe passes the usual check (0.95 on raw scans), rerunning its recipe gave very different accuracies on the same outputs (Table~\ref{tab:main}); for CycleGAN, anywhere between 0.13 and 0.59. Accuracy on raw images did not reveal this: one probe with raw BA 0.98 gave CycleGAN 0.20. Converged probes varied less across seeds, and on every output except NeuroCombat and HACA3 all five scored above the frozen probe's 95\% interval; for aggressive StarGAN, 0.65--0.81 instead of 0.21. The frozen probe's verdict is thus one draw from an unstable process. Because the converged recipe changes several settings at once, it shows that the verdict is sensitive to training, not why.

\textbf{A blur fools the frozen probe.} The Gaussian blur kept PSNR higher than CycleGAN or diffusion (Fig.~\ref{fig:qualitative}) and moved intensities away from NYU, yet it brought the frozen probe down to 0.07, the level of a probe trained on shuffled labels. The probe labelled almost every blurred image as one site, Yale, and none as NYU, whereas most of its errors on CycleGAN, aggressive StarGAN and diffusion outputs pointed to NYU. Checking where errors point, as PRISM does, therefore exposes the blur; as HACA3 shows below, it cannot reveal site information that remains in the outputs.

\textbf{Site remains decodable after harmonization.} Raw-trained slice probes dropped from 0.92 on raw slices to 0.33--0.45 on histogram-matched, CycleGAN and diffusion outputs, a drop that would usually be read as successful harmonization~\cite{galada2025prism}. Probes trained on the outputs themselves, however, recovered the site at 0.87--0.91 (Fig.~\ref{fig:verdicts}a). Head shape alone identifies sites, so this gap could reflect head shape rather than intensities. We therefore repeated the comparison on brain-only slices, with skull and scalp removed: probes trained on the outputs still clearly outperformed raw-trained ones (Fig.~\ref{fig:verdicts}b), and so did probes that see only intensity histograms. The only method that hid the site from such histograms was histogram matching, and only on whole heads, which it matches by construction.

\textbf{HACA3 shows the same gap.} Raw-trained probes scored low on its outputs (0.25--0.29), and most frozen-probe errors pointed to NYU. Yet probes trained on its outputs recovered the site at 0.75, and its intensity histograms alone at 0.71. Within the brain, image probes did no better than brain shape alone, but intensity histograms still recovered the site (0.48 vs.\ 0.10 when trained on raw images). Part of HACA3's effect came from its preprocessing: bias correction and registration alone lowered the frozen probe from 0.95 to 0.61, and HACA3 took it to 0.25. Five-fold cross-validation over all 928 source subjects gave the same picture (0.25 for raw-trained and 0.77 for HACA3-trained probes).

\textbf{High PSNR can mean poor alignment.} DLEST-style 1000 had the highest PSNR (28.4 dB) but moved only a quarter as far toward NYU as CycleGAN ($\Delta W_{\mathrm{NYU}}$ of $-0.009$ vs.\ $-0.036$). Diffusion barely moved toward NYU ($-0.001$), and its two draws differed more from each other than from the input (median foreground mean absolute error 0.085 vs.\ 0.075), so one draw does not represent the method.

\section{Discussion}
\label{sec:discussion}
A single site-probe number is not a harmonization verdict: it changed by up to 0.46 between training seeds, was lowest for a blur, and missed the site information that HACA3's intensities kept. We therefore recommend reporting (1) several well-trained probes, with their recipe, their accuracy on raw images, a shuffled-label baseline and similarity to the input; (2) probes trained on the harmonized outputs, with controls for head shape and intensity, keeping in mind that high accuracy shows remaining site information, not by itself failed harmonization; (3) alignment with the target separately from the amount of change; and (4) uncertainty over subjects and over generator samples.

\textbf{Limitations.} We studied one cohort, one target site and one slice per subject, with 2--15 test subjects per source site. Without ground truth, PSNR and XCorr measure agreement with the input, not correctness. Site is entangled with demographics, anatomy and acquisition geometry; our controls assess these contributions but cannot separate scanner effects from biological ones. Apart from HACA3, the methods are our own untuned implementations, so this is a focused counterexample rather than a survey.

\section{Compliance with Ethical Standards}
\label{sec:ethics}
This research study was conducted retrospectively using human subject data made available in open access by ABIDE~I~\cite{dimartino2014abide}, whose sites shared the fully anonymized data with local ethics approval or an explicit waiver. Ethical approval was not required as confirmed by the license accompanying the publicly available data.

\section{Acknowledgments}
\label{sec:acknowledgments}
ABIDE~I was supported by NIMH grants K23MH087770 and R03MH096321 and the Stavros Niarchos and Leon Levy Foundations. This work also benefited from the support of the project HoliBrain of the French National Research Agency (ANR-23-CE45-0020-01). Finally, this study received financial support from the French government in the framework of the PEPR StratifyAging, the PEPR Prodrom-ND and the IHU VBHI (ANR-23-IAHU-0001). The authors declare no conflicts of interest. OpenAI Codex and Anthropic Claude helped write evaluation and plotting code and edit the text; the authors wrote most of the text, designed the study, reviewed the code, verified all numbers and take full responsibility.

\let\oldthebibliography\thebibliography
\renewcommand{\thebibliography}[1]{\oldthebibliography{#1}%
  \setlength{\itemsep}{0pt}\setlength{\parskip}{0pt}\setlength{\parsep}{0pt}}
{\footnotesize
\bibliographystyle{IEEEbib}
\bibliography{refs}}

\end{document}